\documentclass{ifacconf}

\usepackage{graphicx}      
\usepackage{natbib}        
\usepackage{color}
\usepackage{comment}
\usepackage{amsmath,amssymb,amsfonts}
\usepackage{algorithm}
\usepackage{algpseudocode}
\usepackage{graphicx}
\usepackage{textcomp}

\newtheorem{remark}{Remark}

\usepackage{mathtools, cuted}
 \usepackage{relsize}
\usepackage{bbm}
\usepackage{multicol}

\def\CD#1{\textcolor{blue}{#1}}
\def\MM#1{\textcolor{magenta}{#1}}

\makeatletter
\let\old@ssect\@ssect 
\makeatother

\usepackage{natbib}
\usepackage{hyperref}

\makeatletter
\def\@ssect#1#2#3#4#5#6{%
  \NR@gettitle{#6}
  \old@ssect{#1}{#2}{#3}{#4}{#5}{#6}
}
\makeatother

\begin{document}
\begin{frontmatter}

\title{One-shot backpropagation for multi-step prediction in physics-based system identification -- EXTENDED VERSION}


\author[DET,CNR]{Cesare Donati}
\author[CNR]{Martina Mammarella} 
\author[CNR]{Fabrizio Dabbene}
\author[DET]{Carlo Novara}
\author[PSU]{Constantino Lagoa}

\address[DET]{DET, Politecnico di Torino, Turin, Italy (e-mail: cesare.donati, carlo.novara@polito.it)}
\address[CNR]{CNR-IEIIT, Turin, Italy (e-mail: martina.mammarella, fabrizio.dabbene@cnr.it)}
\address[PSU]{EECS, The Pennsylvania State University, University Park, PA, USA (e-mail: cml18@psu.edu)}

\begin{abstract}
%
The aim of this paper is to present a novel physics-based framework for the identification of dynamical systems, in which the physical and structural insights are reflected directly into a backpropagation-based learning algorithm. The main result is a method to compute in \textit{closed form} the gradient of a \textit{multi-step} loss function, while enforcing physical properties and constraints.
The derived algorithm has been exploited to identify the unknown inertia matrix of a space debris, and the results show the reliability of the method in capturing the physical adherence of the estimated parameters.
\end{abstract}


\end{frontmatter}

\section{Introduction}\label{sec:intro}
In real-world applications, systems of interest are often not precisely known, and physically-consistent approximating models are challenging to identify.
This is especially true in modern problems, which often involve complex, nonlinear, and possibly interconnected systems \citep{ljung2011networked_sys}. 
Moreover, incorporating physical insights while preserving simulation accuracy is not trivial, demanding a fusion between theoretical understanding and computational accuracy.

To overcome these issues, solutions based on the minimization of a multi-step loss function have been proposed 
\citep{mohajerin2019multistep}, providing satisfactory performance in simulation at the expense of a high computational effort and involving, in general, solution of hard non-convex problems.  

Recently, a new model class has become the subject of relevant research activities,  the so-called  \textit{physics-informed neural networks} (PINNs) \citep{karniadakis2021pbnn}. These kinds of NNs are positioned between grey-box and black-box models, and allow to incorporate the available physical information, either by introducing a physics-based loss function \citep{GOKHALE_PINN_loss}, or directly modifying the structure of the model ensuring a consistent physical correlation between input and output \citep{JonesC_PCNN}. PINN techniques have been gaining large interest for their capability of handling the main challenges posed by modern system identification. However, in PINNs usually the NN weights lack of physical interpretability.

Motivated by the previous considerations, in this paper we propose a novel identification framework, which places itself at the intersection of classical grey-box identification, where often nonlinear phenomena are ignored or simplified, and modern PINN methods, where a black-box model is embedded with prior knowledge of the system's physics \citep{NghiemTruongX_PIML}, aiming to exploit the best features of these approaches. The method is based on a (possibly partial) knowledge of the physical description of a nonlinear system, which is used for the definition of a NN-like structure as a substitute for the system dynamical multi-step model. 
Relying on such a model structure, we develop a gradient-based identification algorithm, exploiting the well-known backpropagation method, typically used for classical NN training. 

The philosophy is similar to classical backpropagation, where we leverage the specific characteristics of our problem. First, we enforce the weights to be the same at \textit{each time step} (i.e., in each layer) along the prediction horizon, since they have the same physical interpretation and being the system \textit{time-invariant}.
Second, in our proposed architecture the ``activation functions'' are fixed using the physical dynamics $f$ in \textit{each layer}. Consequently, the \textit{weights} have an \textit{explainable} and \textit{interpretable} meaning, representing the \textit{physical} parameters of the system $\mathcal{S}$ to be identified. Similarly, in \citep{abbasi2022physical} the authors introduce the concept of \textit{physical activation functions} (PAFs), where the mathematical expression of the activation function is inherited from the physical laws of the investigated phenomena. However, these PAFs are applied only one \textit{hidden layer}, and combined with other general activation functions, e.g., sigmoids. 

This formulation allows the definition of an \textit{analytical} and \textit{recursive} computation of the gradient, that exploits all the available physics-based constraints on the system states and parameters and, if any, the system structural information. 
In a conventional neural network, where no incorporation of physics is enforced within the structure, and various activation functions are distributed across layers, obtaining an analytical formulation would have been unfeasible.
The \textit{generality} of the underlying structure allows us to deal with real-world situations where the system to identify may be partly inherited from the physics and partly unknown, and the values of some parameters may be available, while others need to be identified. Moreover, the proposed approach allows to reflect the physical characteristics of the system behavior through the introduction of specific penalty terms in the cost function \citep{ferraritrecate2022physicalconstraints, medina2023active}, ensuring models adherence to fundamental physics principles.

The remainder of the paper is structured as follows. In Section~\ref{sec:framework}, we define the considered framework, introducing the main features of the considered system dynamics and of the estimation model. The analytic computation of the gradient is detailed in Section~\ref{sec:cfgrad}, together with the approach used to enforce possible physics-based constraints based on prior knowledge of the system. Simulation results obtained with the proposed approach are discussed in Section~\ref{sec:results}. Main conclusion are drawn in Section~\ref{sec:concl}.

\subsubsection{Notation}
Given a vector $v$, we denote by $\mathbf{v}_{1:T}\doteq\{{v_k}\}_{k=1}^{T}$  the set of vectors $\{v_1,\ldots,v_T\}$.
Given integers $a\leq b$, we denote by $[a,b]$ the set of integers $\{a,\ldots,b\}$.
The Jacobian matrix of $\alpha_k$ with respect to $\beta_k$ is denoted as $\mathcal{J}^{\alpha\!/\!\beta}_k \in \mathbb R^{n_\alpha\times n_\beta}$
i.e. $\frac{\partial\alpha_k}{\partial\beta_k}$. Similarly, $\mathcal{J}^{\alpha\!/\!\alpha}_k \in \mathbb R^{n_\alpha\times n_\alpha}$ is the Jacobian matrix of $\alpha_k$ with respect to $\alpha_{k-1}$, i.e. $\frac{\partial\alpha_k}{\partial\alpha_{k-1}}$. 

\section{Framework definition}\label{sec:framework}
\subsection{Problem setup}
We consider a dynamical system $\mathcal{\widetilde S}$ and a model $\mathcal{S}$, sufficiently expressive to describe $\mathcal{\widetilde S}$. The model $\mathcal{S}$ is assumed to be  nonlinear, time-invariant, and possibly composed by interconnected subsystems. The model is physics-based, i.e.\ it is defined by means of difference equations capturing the physical interaction between variables, that is it takes the form
\begin{equation}
\begin{aligned}
\mathcal{S}:\,\,\,\,\,\,\,\,&x_{k+1} = f\left(x_k, u_k, \theta\right),\\
    &z_{k} = g\left({x}_{k}\right),
\end{aligned}
\label{eqn:x}
\end{equation}
where $x\in\mathbb{R}^{n_{x}}$ is the state vector, ${u}\in\mathbb{R}^{n_u}$ is the (external) input vector to $\mathcal{S}$, and $z\in\mathbb{R}^{n_{z}}$ is the observation vector. The functions $f(x,u,\theta)$ and $g(x)$ are known, and represent the dynamical laws and the observation function respectively. They are assumed to be nonlinear, time-invariant, and at least $C^1$ differentiable. The goal is to identify both physical parameters $\theta\in\mathbb{R}^{n_\theta}$ and initial condition $x_0\in\mathbb{R}^{n_{x}}$ starting from measured input-output sequences, leading to an estimation model $\mathcal{\widehat S}$ of $\mathcal{S}$ of the form
\begin{equation}
\begin{aligned}
    \mathcal{\widehat S}:\,\,\,\,\,\,\,\,&\hat{{x}}_{k+1} =  f(\hat{ {x}}_{k},  {u}_{k}, \hat \theta),\;\\
    &\hat {{z}}_{k} = {g}(\hat{{x}}_{k}),
    \label{eqn:xhat}
    \end{aligned}
\end{equation}
where $\hat{x}_{k}$ and $\hat{z}_{k}$ are the estimated state and output at time $k$, respectively.

We assume we have available a $T$-step measured, input sequence $\mathbf{\widetilde u}_{0:T-1}$ and the corresponding $T$ collected observations
$\mathbf{\widetilde z}_{0:T-1}$. 
The objective is to estimate the optimal values of the parameters $\hat\theta^\star$ and initial condition $\hat x_0^\star$ over the horizon $T$ such that $\mathcal{\widehat S}$ is the best approximation of $\mathcal{\widetilde S}$, given the underlying physical structure $\mathcal{S}$ and the measured data $\{\mathbf{\widetilde u}_{0:T-1},\mathbf{\widetilde z}_{0:T-1}\}$\footnote{The proposed algorithm can be adapted to the case of multiple trajectories with the same length $T$.}. To this aim, a criterion for assessing the \textit{closeness} between $\mathcal{\widetilde S}$ and $\mathcal{S}$ is defined, in terms of a \textit{{loss function}}. Then, as usual,  the identification problem simply recasts as an \textit{optimization problem}.

First, given the output predictions $\hat z$ and the true measurements $\widetilde z$, we define the prediction error at time $k$ as
\begin{equation}
    e_k \doteq \hat z_k - \widetilde z_k,
    \label{eqn:error}
\end{equation}
and the local loss at time $k$ defined by the weighted norm of the error,
\begin{equation}
     \mathcal{L}(e_k, {\theta})\doteq \frac{1}{T}\| e_k \|_\mathcal{Q}^2 \doteq \frac{1}{T}e_k^\top\mathcal{Q}e_k,
\end{equation}
with $\mathcal{Q} \succeq 0$.

In this paper, we consider a \textit{multi-step regression cost} $\mathcal{C}$ as a sum of local losses  over the prediction horizon $T$ as
\begin{equation}
    \mathcal{C}(e_k,\theta) =    \sum_{k=0}^{T-1} \mathcal{L}(e_k, {\theta})\doteq
    \sum_{k=0}^{T-1} \mathcal{L}_k.
    \label{eqn:cost_general}
\end{equation}
 Then, we can define our nonlinear, parametric model identification problem as
\begin{equation}
    (\hat{\theta},\hat{{x}}_0)\doteq \arg \min_{ \theta, {{x}}_0}\,\, \mathcal{C}(e_k,\theta),
    \label{eqn:opt_prob}
\end{equation}
in which we want to minimize the mean squared error over sampled measurements to obtain an estimate of $\theta$ and $x_0$. 

\subsection{{Multi-step dynamics propagation}} \label{subsec-PINN}
Given the dynamical model $\mathcal{S}$, it is possible to propagate each state variable $x_i$, $i \in [1, n_x]$ over a desired horizon $T$ simply applying the model $\mathcal{S}$ \textit{recursively}, i.e.,
\begin{equation}
    x_{i,k+1} = f_i\left(x_k, u_k, \theta\right), k\in[0,T].
    \label{eqn:prop}
\end{equation}

\begin{figure}[!ht]
    \centering
    \includegraphics[trim=2.5cm 0 0 0cm, clip, width = 0.63\columnwidth]{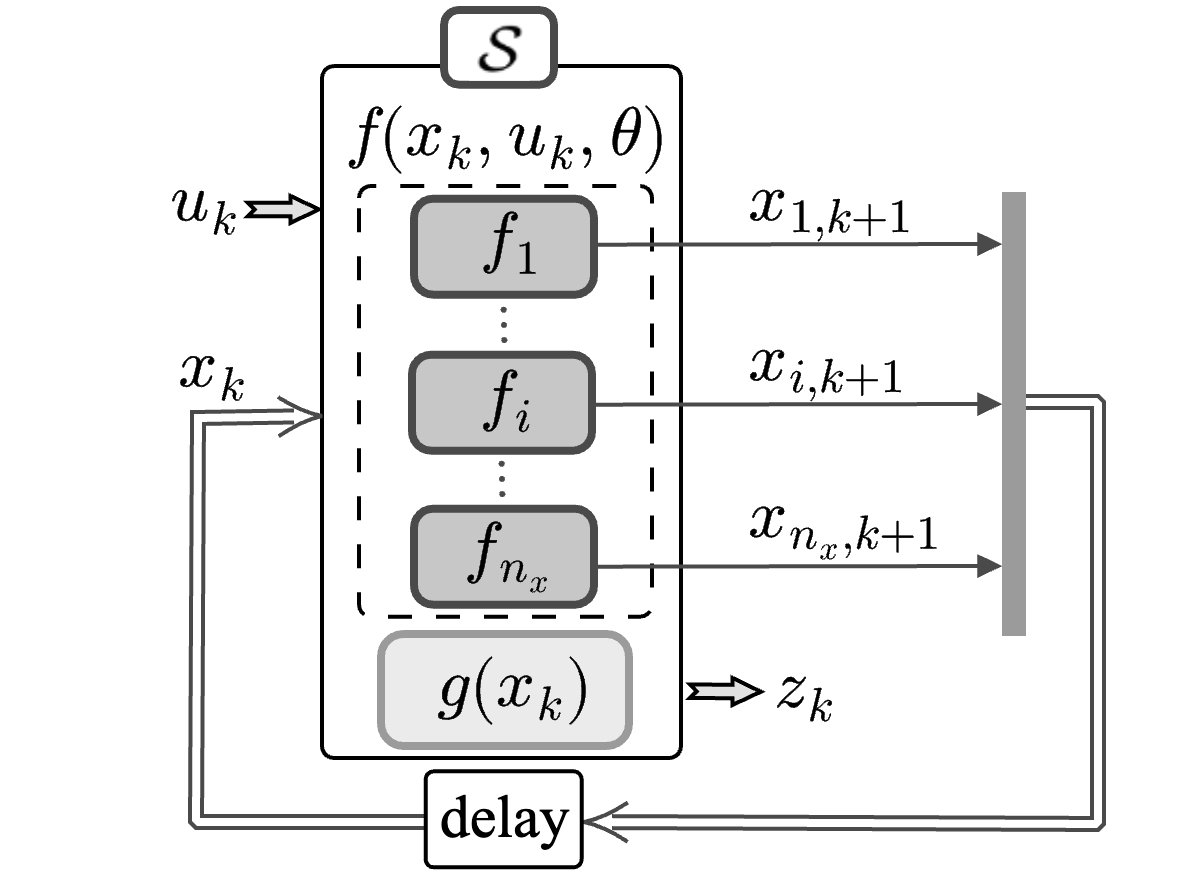}
    \caption{Recursive representation of a dynamical system.}
    \label{fig:FIG1}
\end{figure}
The model can be depicted as in Fig.~\ref{fig:FIG1}, where the
recursion is captured by the delay block. Clearly, this can also be represented opening the output loop $T$ steps ahead from the initial time $k=0$. 

\begin{figure}[!ht]
    \centering
    \includegraphics[width = 0.9\columnwidth]{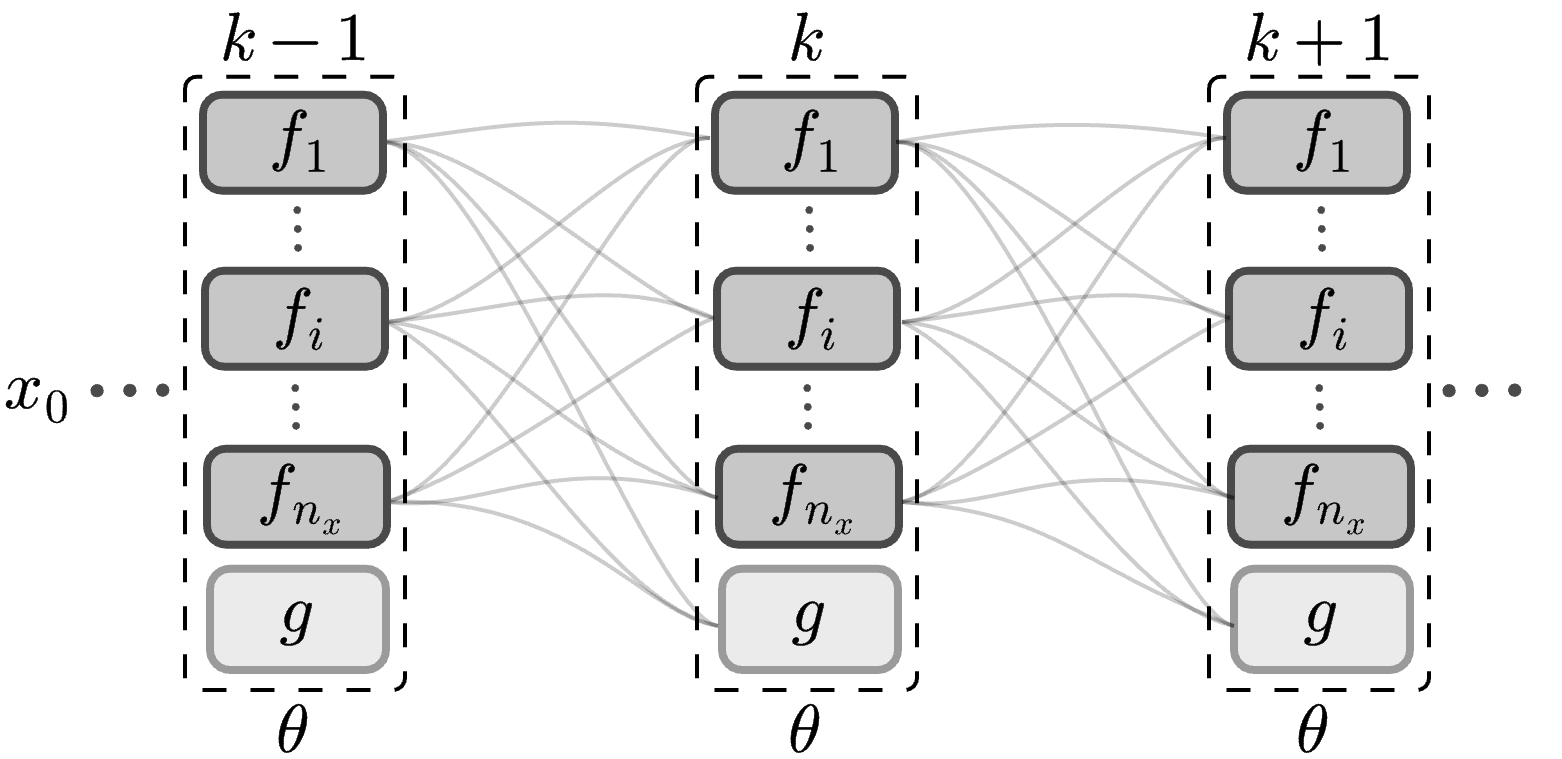}
    \caption{Multi-step system identification structure.}
    \label{fig:NNlike}
\end{figure}

We observe that what we obtain closely resembles the well-known structure of neural networks, as shown in Fig.~\ref{fig:NNlike}. 
Indeed, each time step $k$ can be seen as a ``\textit{layer}'' composed by $n_x$ ``\textit{neurons}'', and the interconnection links between layers and neurons, are activated or deactivated according to the system dynamical structure defined in $\mathcal{S}$. In particular, if $x_{i,k+1}$ does not depend on $x_{j,k}$, the corresponding link is null. This allows to envision the model $\mathcal{S}$ as a neural network graph and, consequently, the ``weights" of the network are the interpretable, physical parameters of the system.

Since the overall objective function in \eqref{eqn:opt_prob} is (in general) \textit{non-convex}, due to the nonlinearity  in $\theta$ and $x_k$ of $f(x_k,u_k,\theta)$ and $g(x_k)$ \eqref{eqn:x}, we rely on gradient-based algorithms \citep{sun2019survey} to address the optimization problem, aiming to reach some (local) minima and eventually compute a (sub)optimal estimation of $\theta$ and $x_0$. 

We observe that, inspired by the approach typically adopted for neural network graphs \citep{pearlmutter1995gradient}, we can exploit a classical backpropagation scheme to analytically compute the gradient of the loss function, thanks to the structure of the physics-based model $\mathcal{S}$. 
However, as it will be clarified in Section \ref{sec:cfgrad}, differently from neural network backpropagation, the scheme in Fig.~\ref{fig:NNlike} presents the \textit{same weigths} $\theta$ and the \textit{same functions} in all layers. This crucial feature allows to derive a useful closed form of the gradient of $\mathcal{C}(e_k,\theta)$ with respect to  $\theta$ and $x_0$, i.e., $ \nabla \mathcal{C} = \left[ \nabla_\theta \mathcal{C},\;\nabla_{x_0} \mathcal{C} \right]$.
Once these gradients are computed, it is possible to apply a gradient-based algorithm to solve the optimization problem \eqref{eqn:cost_general}, such that the estimate of $\theta$ and $x_0$ are updated at each epoch $\ell$. For instance, if a classical gradient descent method is applied, we would have
\begin{align}
\hat\theta^{(\ell+1)}&=\hat\theta^{(\ell)}-\eta_{\theta} \nabla_{\theta} \mathcal{C}^{(\ell)}\\
\hat x_0^{(\ell+1)}&=\hat x_0^{(\ell)}-\eta_{x_0} \nabla_{x_0} \mathcal{C}^{(\ell)}
\end{align}
with learning rates $\eta_{\theta},\eta_{{x}_0}$.
In this paper, we select the ADAM  first-order method \citep{kingma2017adam} with decay rates $\beta_1,\beta_2$.

The whole procedure is presented in Algorithm \ref{alg:algGD}. At epoch $\ell$, we first propagate the system with initial conditions $\hat x_0^{(\ell)}$ and parameters $\hat\theta^{(\ell)}$ through the network layer-by-layer (i.e.\ along the horizon $T$). Then,  we evaluate the gradient based on the computed predictions, and accordingly, we update  the weights, i.e., $\hat\theta^{(\ell)}$ and $\hat x_0^{(\ell)}$. 
This process repeats over $\ell$ until at least one of the following conditions is satisfied: (a) the maximum number of epochs, i.e.\ $E_{max}$, is reached; (b) the structure converges to a (possibly local) minimum of the loss function, or below a given threshold $\varepsilon$; (c) the magnitude of the gradient is lower than a given minimum step size $\delta$.

\begin{algorithm}
\caption{{Backpropagation-based Identification}}\label{alg:algGD}
\begin{algorithmic}[1]
\State Given $T$ input-output observations $\{\mathbf{\widetilde u}_{0:T-1},\mathbf{\widetilde z}_{0:T-1}\}$, choose $\eta_\theta$, $\eta_{x_0}$, $\beta_1$, $\beta_2$, $E_{max}$, $\varepsilon$, and $\delta$.
\State {Initialize} $\ell = 0$ and $\hat{x}^{(0)}_0$, $\hat{\theta}^{(0)}_0$.
\While{ $\ell \leq E_{max}$ \textbf{and} $\mathcal{C}^{(\ell)} \geq \varepsilon$ \textbf{and} $\|\nabla\mathcal{C}\|_2 \geq \delta$ }
\State 
Simulate \eqref{eqn:xhat} for $k\in[0,T-1]$ using $\hat \theta^{(\ell)}$, $\hat x_0^{(\ell)}$
to obtain $\mathbf{\hat x}^{(\ell)}_{1:T}, \mathbf{\hat z}^{(\ell)}_{0:T-1}$.
\State Compute $\mathbf{e}^{(\ell)}_{0:T-1}$ \eqref{eqn:error} and $\mathcal{C}^{(\ell)}$ \eqref{eqn:cost_general}.
\State Compute $\nabla_\theta \mathcal{C}^{(\ell)}$ \eqref{eqn:cfg} and $\nabla_{x_0} \mathcal{C}^{(\ell)}$ \eqref{eqn:cfg0}. 
\State Update the weights using ADAM, i.e., 
$$\hat\theta^{(\ell+1)}=\text{ADAM}(\hat\theta^{(\ell)},\eta_{\theta}, \beta_1, \beta_2, \nabla_{\theta}\mathcal{C}^{(\ell)}),$$
$$\hat{x}_0^{(\ell+1)}=\text{ADAM}(\hat{x}_0^{(\ell)},\eta_{{x}_0}, \beta_1, \beta_2, \nabla_{{x}_0}\mathcal{C}^{(\ell)}).$$
\State $\ell \gets \ell + 1$.
\EndWhile
\State Return $\hat\theta^\star=\hat\theta^{(\ell)}$ and $\hat{x}_0^\star=\hat{x}_0^{(\ell)}$
\end{algorithmic}
\end{algorithm}

\section{Closed-form gradient computation}\label{sec:cfgrad}
In this section, we describe the procedure to compute the gradient in closed form relying on the structure of $\mathcal{S}$ and the available measurements. 
In particular, we compute the gradient of the cost function $\mathcal{C}$ with respect to ${\theta}$ and ${x}_0$, i.e., 
$
\nabla_{\theta} \mathcal{C} = \frac{\mathrm{d} \mathcal{C}}{\mathrm{d} \theta}$ and $\nabla_{{x}_0} \mathcal{C} = \frac{\mathrm{d} \mathcal{C}}{\mathrm{d} {x}_0 }
$
as the product of some intermediate partial derivatives that, unlike what happens in standard neural networks, share a common formulation and allow to compute the gradient analytically. 
Hence, at epoch $\ell$, the analytic form of the gradient can be simply \textit{evaluated} at the current value of $\hat\theta^{(\ell)},\hat x_0^{(\ell)}$ and the ensuing predictions, that is
$$
\begin{aligned}
    \nabla_{\theta} \mathcal{C}^{(\ell)}&= G_\theta \left(
    \hat \theta^{(\ell)}, \hat x_0^{(\ell)},\mathbf{\hat x}_{1:T}^{(\ell)}, \mathbf{\hat z}_{0:T-1}^{(\ell)}\right)\\
    \nabla_{{x}_0} \mathcal{C}^{(\ell)} &= G_{x_0} \left(    
    \hat \theta^{(\ell)}, \hat x_0^{(\ell)},
    \mathbf{\hat x}_{1:T}^{(\ell)}, \mathbf{\hat z}_{0:T-1}^{(\ell)}\right).
\end{aligned}
$$
The closed-form expressions for the two gradients are presented in the following sections. In the sequel, for readability, we omit the superscript $(\ell)$ denoting the epochs.

\subsection{Gradient with respect to parameters}

In the proposed framework, we can obtain the closed-form expression of $\nabla_{\theta} \mathcal{C}$ on the measured data $\{\mathbf{\widetilde u}, \mathbf{\widetilde z}\}$ by considering the effect of the (current, in terms of epochs) estimate $\hat\theta$ \textit{for each time step $k$} on the cost~$\mathcal C$. The desired gradient can be obtained as 
\begin{equation}
    \nabla_{\theta} \mathcal{C} = \sum_{k=1}^{T-1} \left. \frac{\mathrm{d} \mathcal{C}}{\mathrm{d} \theta} \right|_k,
    \label{eqn:cfgradient_easy}
\end{equation}
where $\left. \frac{\mathrm{d} \mathcal{C}}{\mathrm{d} \theta} \right|_k$ is the effect of $\hat\theta$ on the cost $\mathcal{C}$ at an arbitrary time step $k$ within the prediction horizon $T$, and for each~$k$ we have 
\begin{equation}
    \left. \frac{\mathrm{d} \mathcal{C}}{\mathrm{d} \theta} \right|_k = \left. \frac{\partial \mathcal{C}}{\partial \theta} \right|_{k|k} + \sum_{\tau = k+1}^{T-1} \left. \frac{\mathrm{d} \mathcal{C}}{\mathrm{d}\theta} \right|_{\tau|k}.
    \label{eqn:effectk}
\end{equation}

Indeed, this analysis takes into account both the ``direct" effect of $\hat\theta$ at time $k$ on $\mathcal{L}_k$, i.e., $\left. \frac{\partial \mathcal{C}}{\partial \theta} \right|_{k|k}$, and the ``collateral" effects, i.e., $\sum_{\tau = k+1}^{T-1} \left. \frac{\mathrm{d} \mathcal{C}}{\mathrm{d}\theta} \right|_{\tau|k}$, on the subsequent local losses $\mathcal{L}_\tau$ for all $\tau\in[k+1,T]$, arising from the propagation of the error originated from $\hat \theta$ to the predicted state $\hat x_k$.

For the first term in \eqref{eqn:effectk}, we can apply the chain-rule of differentiation, as typically done in classical backpropagation, and we obtain
\begin{equation}
    \begin{aligned}
        \left.\frac{\partial \mathcal{C}}{\partial \theta}\right|_{k|k} &=  
        {\frac{\partial \mathcal{L}_k}{\partial {\theta}}}+
        \frac{\partial \mathcal{L}_k}{\partial e_k} 
        \frac{\partial e_k}{\partial z_k}
        \frac{\partial z_k}{\partial {x_k}}
        \frac{\partial {x_k}}{\partial {\theta}}\\
        &={\nabla_{\theta} \mathcal{L}_k} + 
        {\nabla_{e} \mathcal{L}_k} 
        {\mathcal{J}^{e\!/\!z}_k}
        {\mathcal{J}^{z\!/\!x}_k}
        {\mathcal{J}^{x\!/\!\theta}_k}.
    \end{aligned}
    \label{eqn:contrib_1}
\end{equation}
Then, for the general term $\left. \frac{\mathrm{d} \mathcal{C}}{\mathrm{d} \theta} \right|_{\tau|k}$, we apply again the chain-rule and we have
\begin{equation}
\begin{aligned}
    \left.\frac{\mathrm{d} \mathcal{C}}{\mathrm{d} \theta}\right|_{\tau|k} &=
    \frac{\partial \mathcal{L}_\tau}{\partial e_\tau} 
    \frac{\partial e_\tau}{\partial  z_\tau}
    \frac{\partial  z_\tau}{\partial { x_\tau}}
    \prod_{c=0}^{\tau-k-1} \frac{\partial { x_{\tau-c}}}{\partial  x_{\tau-c-1}}
    \frac{\partial { x_k}}{\partial { \theta}}\\
    &=
    {\nabla_{e} \mathcal{L}_\tau} 
    {\mathcal{J}^{e\!/\! z}_\tau} 
    {\mathcal{J}^{ z\!/\! x}_\tau} \prod_{c=0}^{\tau-k-1} {\mathcal{J}^{ x\!/\! x}_{\tau - c}}
    {\mathcal{J}^{ x\!/\! \theta}_k},
\end{aligned}
\label{eqn:contrib_2}
\end{equation}
where the chain-multiplication of ${\mathcal{J}^{x\!/\!x}}$ evaluated at different time-steps is exploited to back-propagate the error from $\tau$ to $k$ and compute the exact desired contribution of $\hat \theta$ to $\mathcal{C}$ due to the propagation of $\hat x_k$ from time $k$ to time $\tau$.

Then, let us define the following two quantities, i.e., 
\begin{equation}
\begin{gathered}
    \gamma_k \doteq {\nabla_{\hat \theta} \mathcal{L}_k},\quad \Gamma_k \doteq {\nabla_{e} \mathcal{L}_k}{\mathcal{J}^{e\!/\!z}_k}{\mathcal{J}^{z\!/\!x}_k},
\end{gathered}
\label{eqn:gammadef}
\end{equation}
such that
\begin{equation}
    \begin{aligned}
        \left.\frac{\partial \mathcal{C}}{\partial \theta}\right|_{k|k} &=  
        \gamma_k+ 
        \Gamma_k\mathcal{J}^{x\!/\!\theta}_k,
    \end{aligned}
    \label{eqn:contrib_21}
\end{equation}
\begin{equation}
\begin{aligned}
    \left.\frac{\mathrm{d} \mathcal{C}}{\mathrm{d} \theta}\right|_{\tau|k} &=
    \Gamma_k \prod_{c=0}^{\tau-k-1} {\mathcal{J}^{x\!/\!x}_{\tau - c}}
    {\mathcal{J}^{x\!/\!\theta}_k},
\end{aligned}
\label{eqn:contrib_22}
\end{equation}

and substituting these terms in \eqref{eqn:effectk}, we obtain the closed-form for computing $\nabla_{\theta} \mathcal{C}$ as
\begin{equation}
    \begin{aligned}
    \nabla_{\theta} \mathcal{C} &= \sum_{k=1}^{T-1} \gamma_k
    +\sum_{k=1}^{T-1} \Gamma_k {\mathcal{J}^{x\!/\!\theta}_k}   \\
    &+\sum_{k=1}^{T-1} \sum_{\tau = k+1}^{T-1} \bigg( \Gamma_\tau \prod_{c=0}^{\tau-k-1} {\mathcal{J}^{x\!/\!x}_{\tau-c}} \bigg){\mathcal{J}^{x\!/\!\theta}_k}.
    \end{aligned}
    \label{eqn:cfg}
\end{equation}
\vspace{.3cm}
\begin{remark}
By incorporating the model structure $\mathcal{S}$ directly into the network structure, the backpropagation of errors can be efficiently computed using the chain-multiplication of the same Jacobian matrix ${\mathcal{J}^{x\!/\!x}_k}$. The parametric computation of this Jacobian can be performed once for all, and later evaluated at different time steps. This will allow to reduce the number of partial derivatives to be computed and, consequently, the computational complexity of the proposed approach.
\end{remark}
\subsection{Gradient with respect to initial condition}
Let us now consider the explicit formulation for the gradient with respect to the initial condition
\begin{equation}
    \nabla_{x_0} \mathcal{C} = \sum_{k=1}^{T-1} \left. \frac{\mathrm{d} \mathcal{C}}{\mathrm{d} x_0} \right|_{k|0}.
    \label{eqn:cfgradient0_easy}
\end{equation}
The analytical expression can be derived by considering the effect of $x_0$ on each subsequent prediction $\hat{x}_k$ and, consequently, on the cost $\mathcal C$. In this case, there is no ``direct" effect of $\hat x_0$ on the final cost, but we must account for the ``collateral" effects of $\hat x_0$ on the subsequent local-losses $\mathcal{L}_\tau$ for all $\tau=[1,T]$. These effects arise from the error originating from $\hat x_0$ and propagated throughout the predictions along $T$. Consequently, we obtain 
\begin{equation}
\begin{aligned}
    \left.\frac{\mathrm{d} \mathcal{C}}{\mathrm{d} x_0}\right|_{k|0} &= 
    \frac{\partial \mathcal{L}_k}{\partial e_k} 
    \frac{\partial e_k}{\partial \hat z_k}
    \frac{\partial \hat z_k}{\partial {\hat x_k}}
    \prod_{c=0}^{k-1} \frac{\partial {\hat x_{k-c}}}{\partial \hat x_{k-c-1}}\\
    &=
    {\nabla_{e} \mathcal{L}_k}
    {\mathcal{J}^{e\!/\!z}_k}
    {\mathcal{J}^{z\!/\!x}_k} \prod_{c=0}^{k-1}
    {\mathcal{J}^{x\!/\!x}_{k - c}},
\end{aligned}
\label{eqn:contrib0_2}
\end{equation}
which in compact form can be rewritten as
\begin{equation}
        \nabla_{x_0} \mathcal{C} = \sum_{k=1}^{T-1} 
        \Gamma_k\prod_{c=0}^{k-1} {\mathcal{J}^{x\!/\!x}_{k - c}}.
        \label{eqn:cfg0}
\end{equation}


\subsection{Physics-based constraints}\label{sec:pbconstraints}
To guarantee the coherence among the physics of the phenomena and the estimated parameters, exploiting the physical laws as activation functions is not sufficient. We still need to reflect the specificity of the system behaviour, such as e.g.\ passivity, monotonicity, divergence, symmetry of variables, stability \citep{medina2023active,ferraritrecate2022physicalconstraints}, thus ensuring that the identified models adhere to fundamental laws and are consistent with physical principles. This aspect can be formally embedded into the cost $\mathcal{C}$
as a \textit{penalty term} that introduces physical constraints of the form
$$h(\hat x_k, \theta) \leq 0,\,\,\, \forall k \in [0, T],$$
with $h: \mathbb{R}^{n_x} \times  \mathbb{R}^{n_\theta} \rightarrow \mathbb{R}$ a time-invariant function, (at least) $C^1$ differentiable. Specifically, the general cost $\mathcal{C}$ is modified as follows
\begin{equation}
    \mathcal{C} = \sum_{k=0}^{T-1} \mathcal{L}_k + \lambda h(\hat x_k, \theta),
    \label{eqn:cost_general_phy}
\end{equation}
where $\lambda \in \mathbb R$ controls the relevance of the physical constraint $h(\hat x_k, \theta)$ such that higher is the violation of the physical properties in the predicted states and weights, larger is the associated loss value. Similarly, equality constraints may be enforced by adding a quadratic penalty term in the cost. 

In this context, it is still possible to apply the closed-form formula for the gradient simply introducing a penalty term in the loss function which will be accounted in the gradient computation. 
Therefore, the general formulation of the cost function $\mathcal{C}(e_k,\theta)$ \eqref{eqn:cost_general} is modified in order to incorporate the penalty term and introduce physical constraints directly into the optimization problem. 
The closed-form for gradient computation remains unchanged, with the exception of the definition of $\gamma_k$ and $\Gamma_k$ \eqref{eqn:gammadef}, which is modified as follows
$$
    \begin{aligned}
        \gamma_k \doteq {\nabla_{\hat \theta} \mathcal{L}_k} + \lambda{\nabla_{\theta} h},
    \end{aligned}
$$
$$
    \begin{aligned}
        \Gamma_k \doteq {\nabla_{e} \mathcal{L}_k}{\mathcal{J}^{e\!/\!z}_k}{\mathcal{J}^{z\!/\!x}_k} + \lambda{\nabla_{x} h}.
    \end{aligned}
$$

Deterministic physical constraints exhibit themselves in a wide range of forms from simple
algebraic equations to nonlinear integer-differential equations and inequalities. Thus, it is possible to enforce a large variety of physics-based constraints through a sharp customization of $h(\hat x_k, \theta)$. 

\subsection{Physics-based penalty term examples}
\subsubsection{Energy conservation}
Let us consider the identification of a mechanical system. One possibility is to introduce a penalty term to ensure that the total energy remains constant throughout the identification process.
In this scenario, the physics-based penalty can be defined as
$$
h(\hat x_k, \theta) \doteq (E(\hat x_k) - E_0)^2
$$
where $E(\hat x_k)$ represents the total energy based on the system's states at time $k$, and $E_0$ is the reference total energy of the system, which can be computed, for example, based on observations. By minimizing this combined loss function during the system identification process, the identified model is more suited to respect the conservation of energy, making it a more accurate representation of the physical system.
\subsubsection{Physical limits}
In some scenarios, the identified model must ensure that the constraints inherent to the system's physical properties are respected. Let us assume that there exists some physical limits on the state variables, $\overline{x} = [\overline{x}_i],\, i\in [1,n_x],\, \overline{x}_i \in (-\infty, \infty)$, such that 
$$\hat x_{i,k} \leq \overline{x}_i \,\, \forall k.$$
Here, the well-known rectified linear unit can be used, i.e.\
$$\text{ReLU}(\hat{x}_k - \overline{x}) \doteq \max(0, \hat{x}_k - \overline{x}).$$

However, since the ReLU function is non-differentiable at zero and defines a penalty term that only linearly penalizes constraint violations, it is advisable to replace it with a differentiable and more stringent approximation. An \textit{exponential barrier function} can be used to define the physics-based penalty term as follows
$$h(\hat{x}_k, \theta) \doteq \|e^{\alpha(\hat{x}_k - \overline{x})}\|_2^2,$$
where $\alpha > 0 \in \mathbb R$ represents a sharpness parameter.

Consequently, a physical lower bound on the states of the form 
$$\hat x_{i,k} \geq \underline{x}_i \,\, \forall k.$$
can be imposed through the physics-based penalty term
$$h(\hat{x}_k, \theta) \doteq \|e^{\alpha(\underline{x} - \hat{x}_k)}\|_2^2.$$
Here, a special case is the \textit{state non-negativity constraint}, where $\hat x_{i,k} \geq 0 \,\, \forall k$, and $h(\hat{x}_k, \theta)$ becomes
$$h(\hat{x}_k, \theta) \doteq \|e^{-\alpha\hat{x}_k}\|_2^2.$$
This term allows us to check if the state variables violate any physical constraints at each time step, encouraging the system to stay within defined physical limits.
\subsubsection{Convex constraints set in the parameters space}
Similar bounding constraints can be defined to enforce limits on the physical parameters being identified. Thus, the constraint
\begin{equation}
    \theta \in \Theta \doteq \left\{\underline{\theta}_i \leq \hat \theta_i \leq \overline{\theta}_i,\,\, i = [1,n_\theta]\right\},
    \label{eqn:cvx_set}
\end{equation}
can be expressed with the following penalty term
$$h(\hat{x}_k, \theta) \doteq \|e^{\alpha(\hat{\theta} - \overline{\theta})}\|_2^2 + \|e^{\alpha(\underline{\theta} - \hat{\theta})}\|_2^2.$$
Alternatively, the identification algorithm can be enhanced by incorporating a projection step immediately after the parameters update following the gradient computation.
In this context, a projection of the parameters onto the specified convex set defined by \eqref{eqn:cvx_set} can be performed whenever a constraint violation occurs as follows
$$\hat \theta_i  = \min ( \overline{\theta}_i, \hat{\theta}_i),\,\, i = 1,\dots,n_\theta$$
$$\hat \theta_i  = \max ( \underline{\theta}_i, \hat{\theta}_i),\,\, i = 1,\dots,n_\theta $$


\section{Numerical results}\label{sec:results}
The attitude dynamics of the satellite is modeled using standard Euler equations, i.e.,
\begin{equation}
\begin{aligned}
    I\dot{\omega} &= M-\omega\times I\omega, \quad \widetilde{\omega}&=\omega+e_\omega,  
\end{aligned}
    \label{eqn:attitude_dyn}
\end{equation}
where $\omega=[\omega_x,\omega_y,\omega_z]^\top$ is the  angular velocity and $\widetilde{\omega}$ the measured output, $I$ is the satellite inertia tensor, $M$ is the input torque, and $e_\omega$ is the measurement noise. {In the follows, we assume $M \sim \mathcal{N}(10^{-5},\sigma_{M_d})$ with $\sigma_{M_d} = 10^{-7} \frac{\text{rad}}{\text{s}}$}, representing for instance solar radiation pressure, and $e_\omega \sim \mathcal{N} (0,\sigma_{\omega})$ with $\sigma_{\omega} = 10^{-4} {\text{rad}}/{\text{s}}$.\footnote{The noise values, despite appearing rather small, are compatible with the case study selected (i.e., around $10\%$ of the state values).}

Here, the objective is to estimate the optimal value for the satellite diagonal inertia matrix (i.e., the physical parameters $\hat \theta$ are the diagonal elements of $\hat I$) and the initial angular velocity $\hat{\omega}_0$ (i.e., $\hat{{x}}_0$), starting from some tentative values $(I,\omega_0)$ and given collected output samples, applying the proposed approach.
For the validation, we generated a sequence of $T = 50$ data, integrating \eqref{eqn:attitude_dyn} with a sampling time of $0.1$ s. The true systems is initialized with $\omega_{0}=[9.915\cdot10^{-6}, -1.102\cdot 10^{-3}, 1.3179\cdot10^{-5}]^\top$ and $\theta=[0.0403, 0.0404, 0.0080]^\top$. 
%

\begin{remark}
    While the emphasis in this section lies on $\theta$ due to its higher significance in the considered framework, it is important to note that the achieved results were obtained by estimating both $\theta$ and $x_0$.
\end{remark}

\begin{figure}[!ht]
    \centering
    \includegraphics[trim=1cm 0cm 0cm 0cm, clip, width=1\columnwidth]{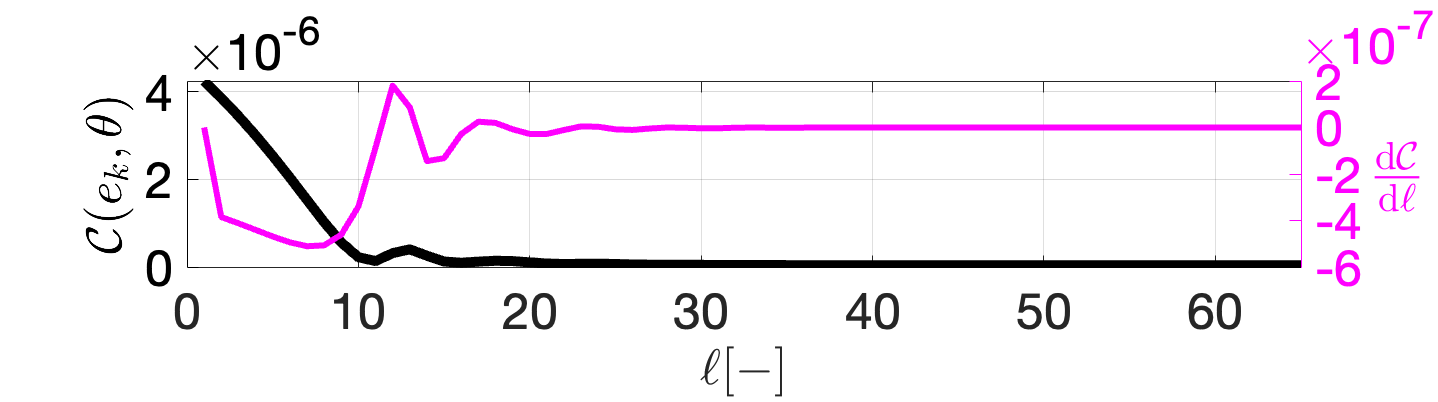}
    \caption{The evolution of $\mathcal{C}$ (black line) and its variation over the iterations (magenta line) for $\ell\in[0,65]$.}
    \label{fig:loss}
\end{figure}

In Fig.~\ref{fig:loss}, we can observe the decreasing, convergent behavior of loss functions over the algorithm iteration epochs $\ell$ on the entire dataset and a similar trend also for the variation of the loss function over $\ell$, i.e., $\mathrm{d}J/\mathrm{d}\ell$.

\begin{figure}[!ht]
    \centering
    \includegraphics[trim=0 0.3cm 0 0cm, clip=true, width=1\columnwidth]{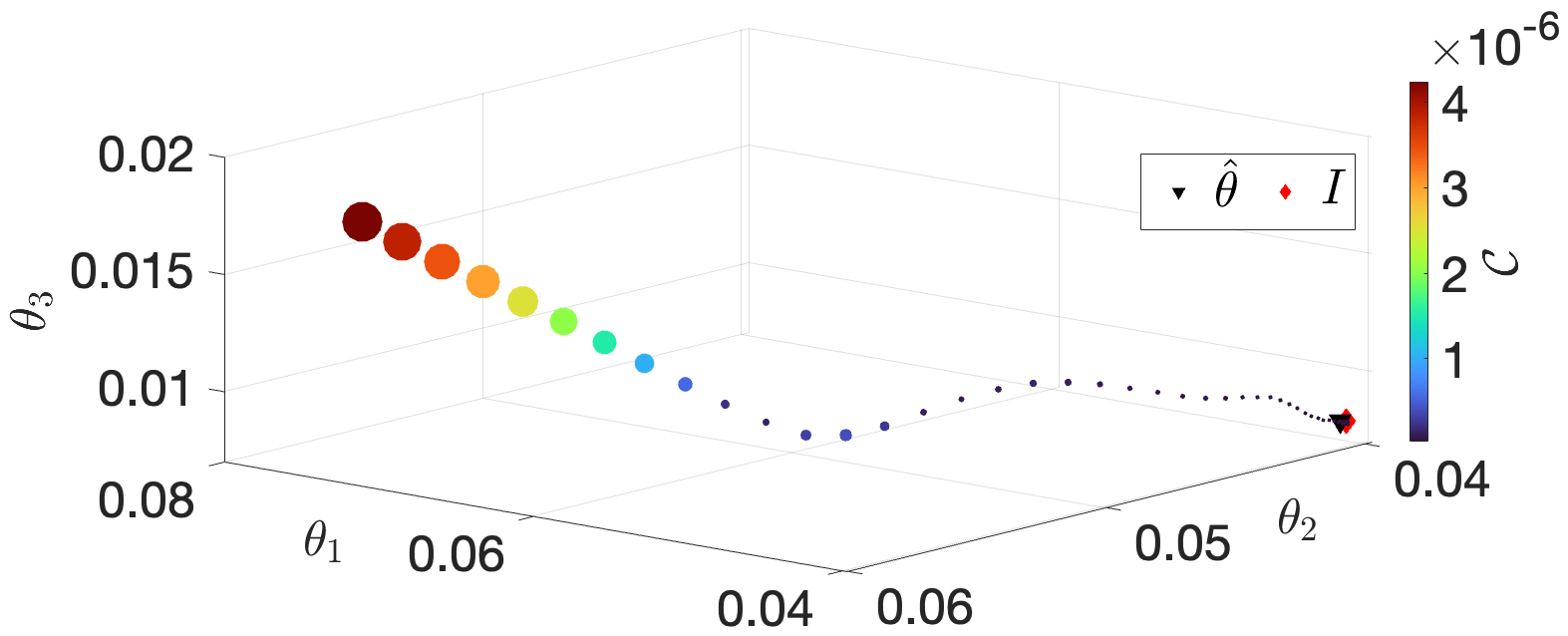}
    \caption{Evolution of $\mathcal{C}$ over the estimation parameter space.}
    \label{fig:J_wrt_th}
\end{figure}

\begin{figure*}
    \centering
    \includegraphics[trim=0cm 0cm 0cm 0cm, clip=true,width=0.66\columnwidth]{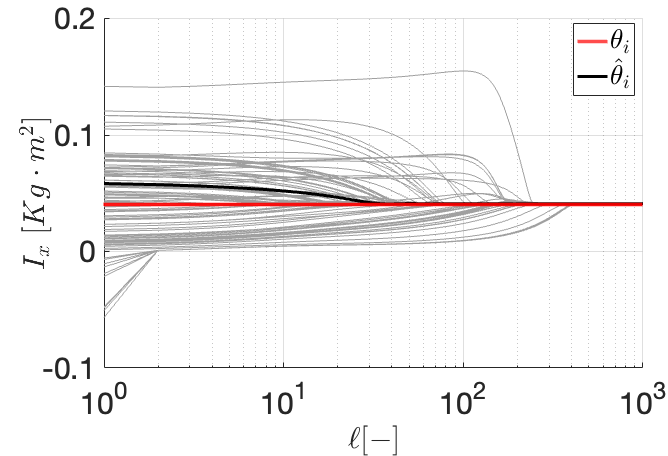}
    \includegraphics[trim=0cm 0cm 0cm 0cm, clip=true,width=0.66\columnwidth]{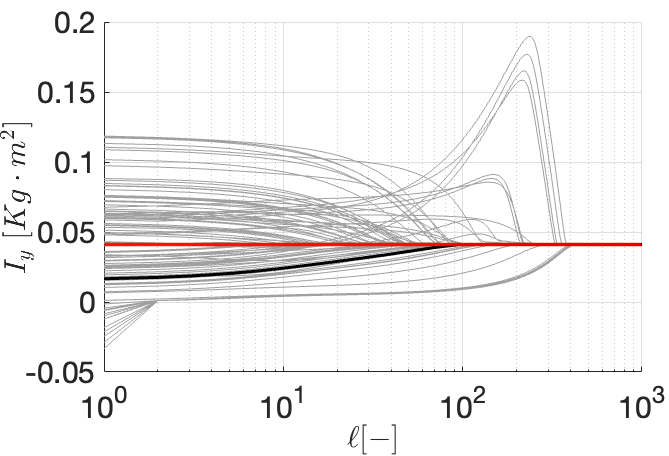}
    \includegraphics[trim=0cm 0cm 0cm 0cm, clip=true,width=0.66\columnwidth]{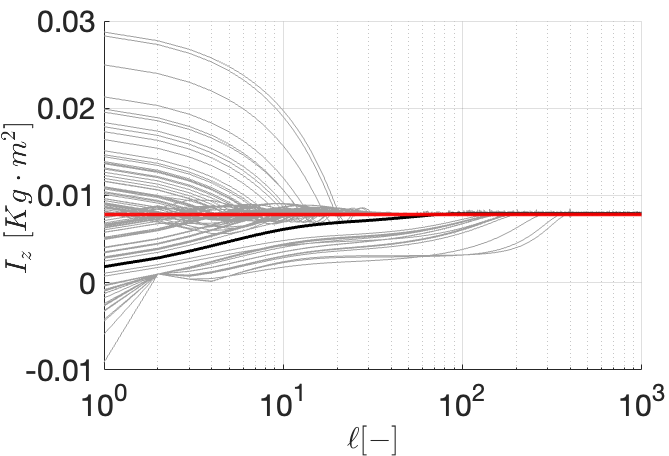}
    \caption{Comparison between estimated parameters $\hat \theta_i$ and real ones $\theta_{i}$.}
    \label{fig:theta_opt}
\end{figure*}

This behavior  is confirmed when represented over the estimated parameter space in Figs.~\ref{fig:J_wrt_th}, \ref{fig:theta_opt}, where we depict the evolution of the estimated parameters with respect to the algorithm epochs $\ell$ for different initial condition of $\hat \theta$.
It is worth noting that the computed gradient might initially move some parameters away from their intended final values (e.g., the peak in the second plot). This temporary shift allows focusing on correcting more crucial parameters first, before eventually re-adjusting the divergent parameter towards convergence.

\begin{figure}[!ht]
    \centering
    \includegraphics[trim=0cm 0cm 0cm 0cm, clip=true,width=1\columnwidth]{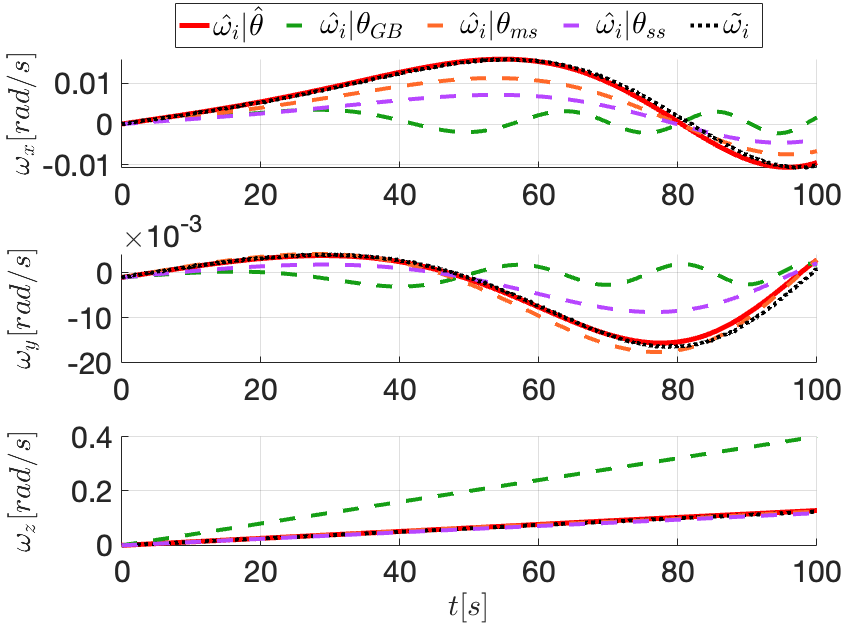}
    \caption{Evolution of $\hat \omega_i(t)$ with different approaches.}
    \label{fig:res06_comp}
\end{figure}

Then, in Fig.~\ref{fig:res06_comp} we compare the performance of the proposed algorithm with respect to three different approaches: (i) a gray-box (GB) model\footnote{We exploited the MATLAB \textit{System identification Toolbox} to implement the GB method, using the \texttt{nlgreyest} function.} (green line), which is fed with the dynamical model in \eqref{eqn:attitude_dyn} and minimizes a single-step prediction error; (ii) a multi-step (ms) model (orange line) and (iii) a single-step (ss) model, both implemented using the same cost function as our approach but different algorithms to compute the gradient, i.e., \texttt{fmincon} function with a \texttt{sqp} setting. \footnote{The comparison with ms is mainly for validation purpose.}
%
%
%
Given the same training dataset, we use all the aforementioned approaches to estimate the physical parameters~$\theta$, and then to propagate the dynamics over a longer simulation horizon (i.e., $t\in[0,100]$), overlapping the results with the real measurements (black line). We can observe that both multi-step approaches are able to properly capture the physics of the system better than the GB and ss. However, we need to emphasize that, due to the inherent instability of the trajectories generated by the nonlinear system~\eqref{eqn:attitude_dyn}, it is expected that also the trajectory estimated using our approach could eventually diverge from the actual one. Indeed, in this context, the goal of multi-step identification is to identify parameters that enable the longest horizon of accurate predictions given a training sequence of $T$ data.

\begin{figure}[h]
    \centering
    \includegraphics[trim= 0cm 0cm 0cm 0cm, clip=true, width=1\columnwidth]{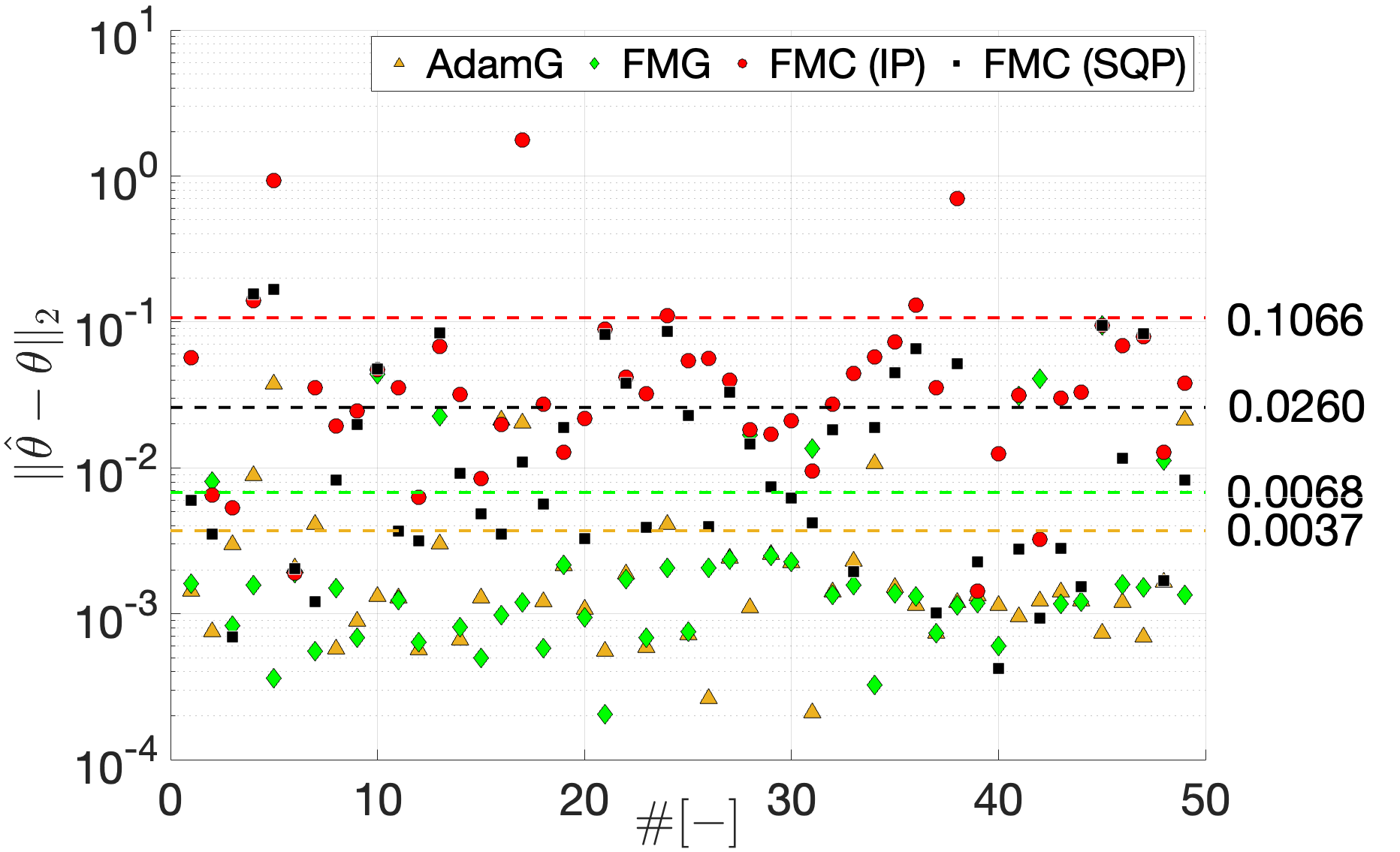}
    \caption{Comparison among four multi-step approaches: 1) \texttt{Adam} with analytic gradient (triangle), 2) \texttt{fmincon} with analytic gradient (diamond), 3) \texttt{ipopt}-\texttt{fmincon} (circle), and 4) \texttt{sqp}-\texttt{fmincon} (square).}
    \label{fig:mscomp}
\end{figure}

Between the two multi-step approaches the main difference resides in the gradient computation, i.e., \textit{analytically computed} in our approach and \textit{numerically approximated} for the standard multi-step approach, and how this affects the estimation algorithm. This is highlighted in Fig.~\ref{fig:mscomp} where we compare three multi-step approaches, sharing the same solver \texttt{fmincon} with $E_{max}=100$, in terms of estimation error $\|\hat \theta-\theta\|_2$. We can notice that using the analytical gradient allows to increase the estimation accuracy by one order of magnitude with respect to \texttt{ipopt} and \texttt{sqp} methods. Moreover, we can observe that, providing the same analytic gradient to two different solvers, i.e. \texttt{fmincon} and \texttt{Adam}, we can achieve an additional improvement with the latter solver.

\begin{figure*}[!ht]
    \centering
    \includegraphics[trim=0cm 0cm 0cm 0cm, clip=true,width=0.63\columnwidth]{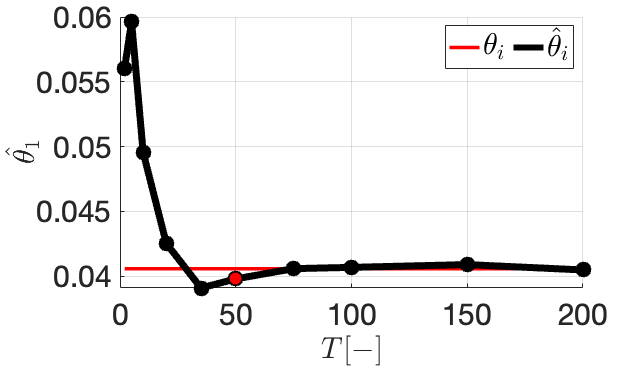}
    \includegraphics[trim=0cm 0cm 0cm 0cm, clip=true,width=0.63\columnwidth]{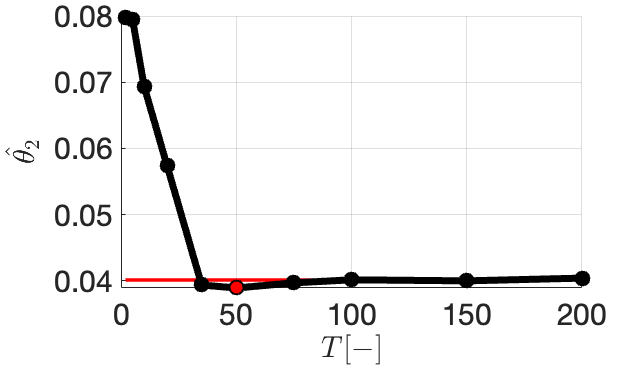}
    \includegraphics[trim=0cm 0cm 0cm 0cm, clip=true,width=0.63\columnwidth]{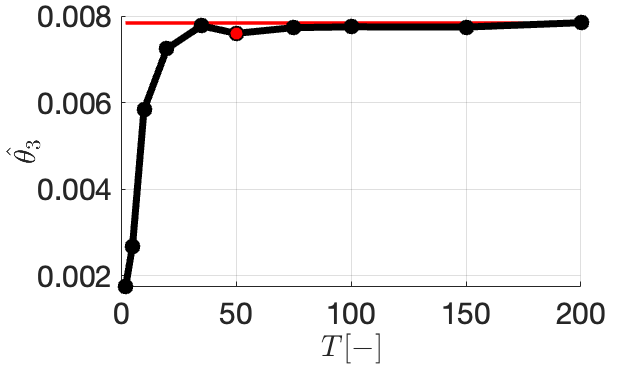}
    \caption{Estimated $\hat \theta_i$ for different prediction horizons $T$.}
    \label{fig:res08_horizon_time1}
\end{figure*}

\begin{figure}[!ht]
    \centering
    \includegraphics[trim=0cm 0cm 0cm 0cm, clip=true,width=1\columnwidth]{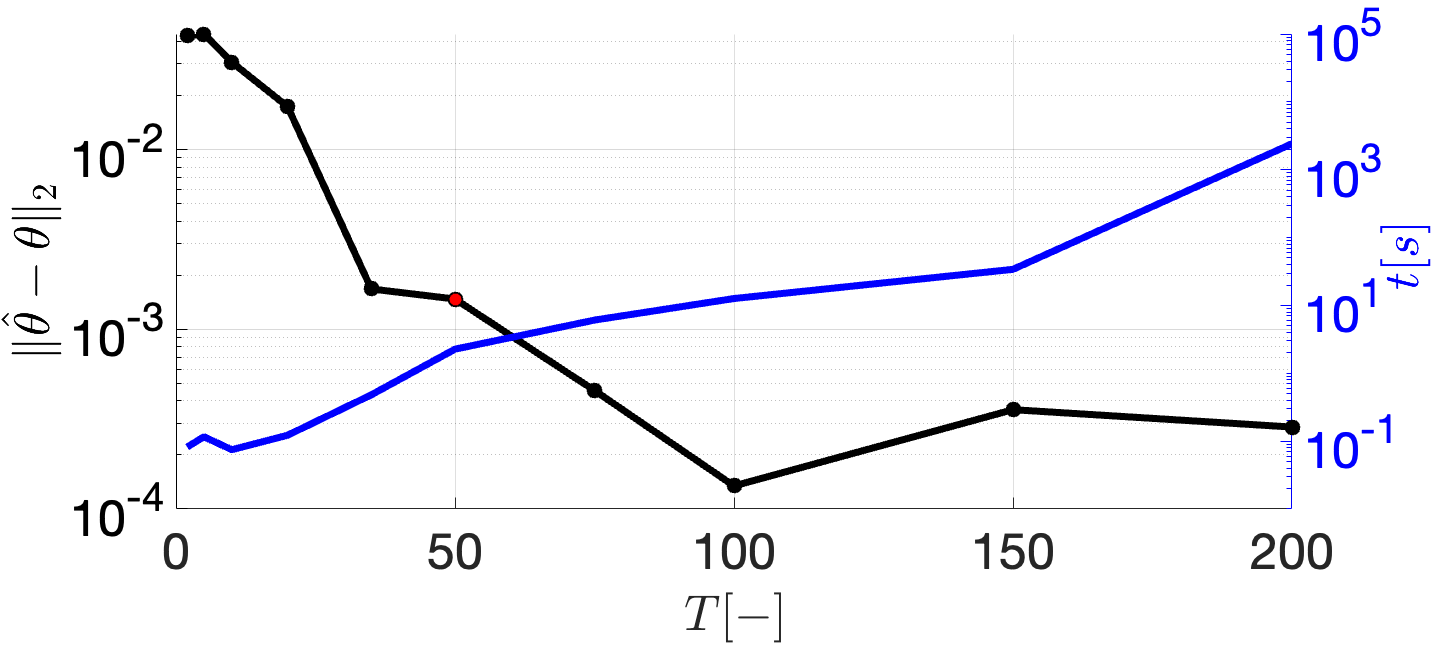}
    \caption{Estimation error for different prediction horizons~$T$.}
    \label{fig:res08_horizon_time}
\end{figure}

The last aspect analyzed is the correlation among the prediction horizon $T$, the quality of the estimated parameters $\hat \theta$ and the computation time for the proposed multi-step identification scheme. To compare the performance with respect to the required time we performed different simulations using different prediction horizons. As shown in Fig.~\ref{fig:res08_horizon_time1},~\ref{fig:res08_horizon_time}, the larger is $T$ (i.e.\ the larger is the number of data used to compute the gradient), the higher the computation time (blue line) required to complete the identification will be. 
Observing the estimation performance, we can select a trade-off horizon between performance improvement and required computation time ($T=50$,~$\hat \theta = [0.0398, 0.0389, 0.0076]^\top$).

\section{Conclusions and future research}\label{sec:concl}

In this work we proposed a general framework for the identification of complex dynamical systems focusing on multi-step prediction accuracy. We presented here the main technical steps, concentrating on the case when a physical description of each subsystem is available.  However, we want to remark that the approach is general, and it can be extended to situations where only partial information on the structure or on the state equations is available. This is the subject of current research. In particular, in the case of partially known equations, the idea is to assume that the model to estimate is given by the sum of two contributions: a term directly modeled according to the (underlying) physics of the system, and another one capturing the unmodeled dynamics.
\bibliography{ifacconf}

\begin{thebibliography}{12}
\providecommand{\natexlab}[1]{#1}
\providecommand{\url}[1]{\texttt{#1}}
\providecommand{\urlprefix}{URL }
\expandafter\ifx\csname urlstyle\endcsname\relax
  \providecommand{\doi}[1]{doi:\discretionary{}{}{}#1}\else
  \providecommand{\doi}{doi:\discretionary{}{}{}\begingroup \urlstyle{rm}\Url}\fi

\bibitem[{Abbasi and Andersen(2022)}]{abbasi2022physical}
Abbasi, J. and Andersen, P.{\O}. (2022).
\newblock {P}hysical {A}ctivation {F}unctions {(PAFs)}: {A}n {A}pproach for {M}ore {E}fficient {I}nduction of {P}hysics into {P}hysics-{I}nformed {N}eural {N}etworks {(PINNs)}.
\newblock \emph{arXiv preprint arXiv:2205.14630}.

\bibitem[{Di~Natale et~al.(2022)Di~Natale, Svetozarevic, Heer, and Jones}]{JonesC_PCNN}
Di~Natale, L., Svetozarevic, B., Heer, P., and Jones, C.N. (2022).
\newblock {Physically consistent neural networks for building thermal modeling: {T}heory and analysis}.
\newblock \emph{Applied Energy}, 325.

\bibitem[{Gokhale et~al.(2022)Gokhale, Claessens, and Develder}]{GOKHALE_PINN_loss}
Gokhale, G., Claessens, B., and Develder, C. (2022).
\newblock {Physics informed neural networks for control oriented thermal modeling of buildings}.
\newblock \emph{Applied Energy}, 314.

\bibitem[{Karniadakis et~al.(2021)Karniadakis, Kevrekidis, Lu, Perdikaris, Wang, and Yang}]{karniadakis2021pbnn}
Karniadakis, G., Kevrekidis, I., Lu, L., Perdikaris, P., Wang, S., and Yang, L. (2021).
\newblock Physics-informed machine learning.
\newblock \emph{Nature Reviews Physics}, 3(6), 422--~440.

\bibitem[{Kingma and Ba(2017)}]{kingma2017adam}
Kingma, D.P. and Ba, J. (2017).
\newblock Adam: A method for stochastic optimization.

\bibitem[{Ljung et~al.(2011)Ljung, Hjalmarsson, and Ohlsson}]{ljung2011networked_sys}
Ljung, L., Hjalmarsson, H., and Ohlsson, H. (2011).
\newblock Four encounters with system identification.
\newblock \emph{European Journal of Control}, 17(5), 449--471.

\bibitem[{Medina and White(2023)}]{medina2023active}
Medina, J. and White, A.D. (2023).
\newblock Active learning in symbolic regression performance with physical constraints.
\newblock \emph{arXiv preprint arXiv:2305.10379}.

\bibitem[{Mohajerin and Waslander(2019)}]{mohajerin2019multistep}
Mohajerin, N. and Waslander, S.L. (2019).
\newblock Multistep prediction of dynamic systems with recurrent neural networks.
\newblock \emph{IEEE Transactions on Neural Networks and Learning Systems}, 30(11), 3370--3383.

\bibitem[{Nghiem et~al.(2023)Nghiem, Drgoňa, Jones, Nagy, Schwan, Dey, Chakrabarty, Di~Cairano, Paulson, Carron, Zeilinger, Shaw~Cortez, and Vrabie}]{NghiemTruongX_PIML}
Nghiem, T.X., Drgoňa, J., Jones, C., Nagy, Z., Schwan, R., Dey, B., Chakrabarty, A., Di~Cairano, S., Paulson, J.A., Carron, A., Zeilinger, M.N., Shaw~Cortez, W., and Vrabie, D.L. (2023).
\newblock Physics-informed machine learning for modeling and control of dynamical systems.
\newblock In \emph{2023 American Control Conference (ACC)}, 3735--3750.

\bibitem[{Pearlmutter(1995)}]{pearlmutter1995gradient}
Pearlmutter, B.A. (1995).
\newblock Gradient calculations for dynamic recurrent neural networks: {A} survey.
\newblock \emph{IEEE Transactions on Neural Networks}, 6(5), 1212--1228.

\bibitem[{Sun et~al.(2019)Sun, Cao, Zhu, and Zhao}]{sun2019survey}
Sun, S., Cao, Z., Zhu, H., and Zhao, J. (2019).
\newblock A survey of optimization methods from a machine learning perspective.
\newblock \emph{IEEE Transactions on Cybernetics}, 50(8), 3668--3681.

\bibitem[{Zakwan et~al.(2022)Zakwan, Di~Natale, Svetozarevic, Heer, Jones, and Trecate}]{ferraritrecate2022physicalconstraints}
Zakwan, M., Di~Natale, L., Svetozarevic, B., Heer, P., Jones, C.N., and Trecate, G.F. (2022).
\newblock Physically consistent neural {ODE}s for learning multi-physics systems.
\newblock \emph{arXiv preprint arXiv:2211.06130}.

\end{thebibliography}
      
\end{document}